\documentclass[12pt,preprint]{aastex}

\usepackage{emulateapj5}

\newcommand{\ieee}{IEEE Trans.~Nucl.~Sci.}
\newcommand{\flux}{{\rm ergs~cm^{-2}s^{-1}}}
\newcommand{\lumi}{{\rm ergs~s^{-1}}}
\newcommand{\phcms}{{\rm photons~cm^{-2}s^{-1}}}
\newcommand{\err}[2]{^{+ #1}_{- #2}}
\newcommand{\fixed}{_{\rm fixed}}
\newcommand{\thomson}{\sigma_{\rm T}}

\newcommand{\GRO}{GRO~J2019+37}
\newcommand{\VCYG}{V444~Cyg}
\newcommand{\NGC}{NGC~6888}
\newcommand{\GEV}{GEV~2035+4213}

\shorttitle{ASCA View of the Supernova Remnant $\gamma$ Cygni}
\shortauthors{Y.Uchiyama et al.}

\begin{document}
\title{\emph{ASCA} View of 
the Supernova Remnant $\gamma$~Cygni (G78.2+2.1) :\\
Bremsstrahlung X-ray Spectrum from Loss-flattened Electron 
Distribution}

\author{
Y.~Uchiyama,\altaffilmark{1,2} 
T.~Takahashi,\altaffilmark{1,2}
F.A.~Aharonian,\altaffilmark{3}
and J.R.~Mattox\altaffilmark{4}}

\altaffiltext{1}{Institute of Space and Astronautical Science,
3-1-1 Yoshinodai, Sagamihara, Kanagawa 229-8510, Japan;
uchiyama@astro.isas.ac.jp}
\altaffiltext{2}{Department of Physics, University of Tokyo,
7-3-1 Hongo, Bunkyo-ku, Tokyo 113-0033, Japan}
\altaffiltext{3}{Max-Planck-Institut f\"{u}r Kernphysik, Postfach 103980,
D-69029 Heidelberg, Germany}
\altaffiltext{4}{Department of Physics \& Astronomy, Francis Marion University,
Florence, SC 29501-0547}

\begin{abstract}
With the \emph{ASCA} X-ray satellite,
we perform spatial and spectral studies of
the shell-type supernova remnant (SNR) $\gamma$~Cygni that is associated with 
the brightest EGRET unidentified source 3EG~J2020+4017. 
At energies below 3~keV  
the bulk of the X-ray flux from the remnant is well 
described by the emission of thermal plasma with 
characteristic temperature $kT_e \simeq $~0.5--0.9~keV.
In addition to this thermal emission, 
we found an extremely hard X-ray component from  
several clumps localized in the northern part of the remnant.
This component, which dominates the X-ray emission from the remnant 
above 4~keV, is described  
by a power-law with a photon index of $\Gamma \simeq $~0.8--1.5.
Both the absolute flux and the spectral shape of 
the nonthermal X-rays cannot be explained 
by the synchrotron or inverse-Compton mechanisms. 
We argue that the unusually hard X-ray spectrum  
can be naturally interpreted in terms of nonthermal bremsstrahlung from  
Coulomb-loss-flattened electron distribution
in dense environs with the gas density about 10 to 100 $\rm cm^{-3}$. 
For given spectrum of the electron population, the ratio of the bremsstrahlung 
X- and $\gamma$-ray fluxes depends on the position of the
``Coulomb break'' in the electron spectrum. 
Formally, the entire high energy $\gamma$-ray flux detected
by EGRET from $\gamma$~Cygni could originate in the hard X-ray regions.
However, it is more likely that the bulk of $\gamma$-rays detected 
by EGRET come from the radio-bright and X-ray dim cloud at 
southeast, where the very dense gas and strong magnetic field would 
illuminate the cloud in the radio and $\gamma$-ray bands, but suppress the 
bremsstrahlung X-ray emission due to the shift of the ``Coulomb break''
in the electron spectrum towards higher energies.  
\end{abstract}


\section{Introduction}
The origin of the unidentified EGRET $\gamma$-ray sources 
in the Galactic plane has been puzzling since their discovery.
The third EGRET catalog \citep{ThirdEGRET} lists 57 unidentified sources
at low Galactic latitude ($|b|\leq 10\arcdeg$).
Rotation-powered pulsars are likely to account for part of 
these GeV $\gamma$-ray sources;
five pulsars have been detected at GeV energies to date.
Another possible origin for some of the unidentified sources
is the emission from accelerated cosmic-rays
at the shock of shell-type supernova remnants (SNRs).
It is reported that the probability to find EGRET unidentified 
sources in the vicinity of shell-type SNRs is significantly high
\citep{Sturner95}.  
Although the relatively young SNRs, being in their Sedov phase, 
are natural sites of high energy $\gamma$-ray production through
electron bremsstrahlung and hadronic interactions,
it has been recognized that in most cases the expected $\gamma$-ray fluxes 
at MeV/GeV energies are too low to be detected by EGRET \citep{Drury94}.  
However, the $\gamma$-ray fluxes can be dramatically enhanced in SNRs 
having dense gas environments, e.g., in large molecular clouds overtaken 
by supernova shells \citep{Aharonian94}.
Remarkably, among the SNRs possibly detected by EGRET are 
the radio-bright and nearby objects $\gamma$~Cygni, IC~443, W44, and 
W28 \citep{Esposito} that are all associated with molecular clouds.
The recent searches for SNRs in the vicinity of some of the 
unidentified EGRET sources revealed new evidence of $\gamma$-ray emission from 
``supernova remnant--molecular cloud'' interacting systems 
\citep{combi98,combi01}.
If gamma-rays of such systems have hadronic origin, 
and the acceleration of protons extends to 10 TeV and beyond,
the TeV $\gamma$-ray emission from these objects should be detectable 
also by current ground-based detectors \citep{Aharonian94}.
At energies below the threshold of production of $\pi^0$-decay
$\gamma$-rays around 70 MeV,
the electron bremsstrahlung remains the only noticeable $\gamma$-ray 
production process.
Therefore, the detection of $\gamma$-rays by EGRET down to 
several tens of MeV implies the existence of low-energy ($< 100$~MeV)
electrons in these objects. 
The best information about higher energy electrons, 
typically between 1 and 10 GeV electrons, is provided 
by the synchrotron radio emission at GHz frequencies.

The accelerated electrons produce also nonthermal 
X-rays through the synchrotron radiation and the bremsstrahlung.
Both channels contain unique information about the nonthermal electron 
populations at extremely high (multi-TeV), and very low (sub-MeV) energies,
respectively. Thus, the X-ray observations may help to reveal the nature
of the unidentified EGRET sources that are associated with SNRs 
by looking for ultrarelativistic and subrelativistic 
electrons, and probing the environments of the remnants.

The $\gamma$~Cygni (G78.2+2.1) SNR has a clear position-correlation with
the brightest unidentified $\gamma$-ray source 3EG~J2020+4017.
It is a nearby (1--2~kpc) shell-type SNR with the radio shell
of $\sim 60\arcmin$ diameter \citep{Higgs77}.
The radio flux density of 340 Jy at 1 GHz ranks it as 
the fourth brightest SNR in the sky at this frequency \citep{Green}.
Almost 60\% of the radio flux comes from southeastern part
which has been known as DR4 \citep{DR}.
The spectral index of the radio spectrum averaged over 
the whole remnant is measured as $\alpha \simeq 0.5$ \citep{Green}.
Its variation across the remnant is as small as
$\Delta\alpha\sim \pm0.15$ \citep{Zhang}.
The EGRET source has the steady flux of 
$F(E>100{\rm~MeV})=(12.6\pm0.7)\times10^{-7}~\phcms$ and 
a best-fit power-law index of $2.07\pm0.05$ \citep{Esposito}.
Prior to the EGRET detection, $\gamma$-ray source 2CG078+2 was detected
in the vicinity of $\gamma$~Cygni
with the \emph{COS-B} satellite \citep{Pollock}.
Despite extensive searches for TeV $\gamma$-ray emission,  
significant excesses have not been detected \citep{Buckley}.
Since a simple extrapolation of the EGRET flux 
exceeds the Whipple upper limit by an order-of-magnitude,
the spectrum must have a cutoff or steepen well below 
the TeV energy \citep{Gaisser98,Buckley}.

Presented here are the results from 
and implications of \emph{ASCA} observations of the $\gamma$~Cygni SNR.
Of particular interest is the discovery of hard X-ray emission
localized to the northern part of $\gamma$~Cygni.
The plan of this paper is as follows.
The \emph{ASCA} observations are briefly summarized in \S\ref{sec:obs}.
We perform the X-ray image analysis in \S\ref{sec:ximage}
and the spectral fits in \S\ref{sec:xspec}.
In \S\ref{subsec:cloud}, 
we interpret the soft X-ray data in terms of 
shock--cloud interaction.
The origin of the hard X-ray emission is discussed in \S\ref{sec:nonthermal}.

\section{Observations}\label{sec:obs}
We performed \emph{ASCA} observations of $\gamma$~Cygni SNR 
twice in 1996 and 1997.
The northern part of the remnant was observed for 40 ks in 1996 May.
Observations were carried out in 1997 May with three pointings toward 
the north, south, and east of $\gamma$~Cygni
with 60, 16, and 12 ks duration, respectively.
The data from two Gas Imaging Spectrometer \citep{Max96}
detectors were acquired in the standard pulse-height mode.
Two Solid-state Imaging Spectrometer \citep{Bur94} detectors 
were operated in 4-CCD mode.
Each detector is coupled to its own telescope with nested 
conical foil mirrors.
Once combined all observations, field of view (FOV) of GIS covers
almost all part of the remnant.

The data from the detectors were screened by standard procedures, 
except for the strict rise-time screening 
in the GIS analysis, because of the mode we chose for the operation.
The observed count rate for the north of $\gamma$~Cygni
was 0.57 (0.59) counts/s
within a central 20\arcmin\ radius in the 0.7--10 keV energy band
with GIS2 (GIS3).
The degradation of the SIS performance due to the accumulated 
radiation damage in orbit is found to be substantial in our data.
Even for the data taken with faint mode, in which 
RDD (Residual Dark Distribution) correction can be applied, 
the spectral resolution of SIS is found to be of no advantage 
to that of GIS.
In addition, the smaller effective area at high 
energies and the narrower FOV for SIS than those for GIS 
do not serve our purpose.
The results presented in this paper are from the GIS data.

\section{Data Analysis and Results}\label{sec:analysis}
\subsection{X-ray Images}\label{sec:ximage}
The \emph{ASCA} GIS data for the four pointings were combined to construct 
the X-ray images in selected energy bands.
We discard the data outside 
the central region of the detector with a radius greater than 20\arcmin .
By utilizing the night-earth observation data set,
we subtract the instrumental background from each pointing data.
Since we did not obtain the rise-time information, the normalization 
of the background image is increased by 30\% as compared with the nominal case.
The background-subtracted images are combined with
corrections for the exposure time and the vignetting effects.
Finally the images are smoothed with 
a function comprised of a narrow core ($\lesssim 1\arcmin$) and a broad wing 
to simulate the point spread function (PSF) of the telescope.
Figure~\ref{fig:ximage}{\it a}, \ref{fig:ximage}{\it b}, and 
\ref{fig:ximage}{\it c} show
the resultant X-ray images in 0.7--1~keV (soft), 1--3~keV (medium),
and 4--7~keV (hard) energy bands, respectively.

We find strikingly different appearance of the SNR in these energy bands.
In the soft band image, clumpy sources appear at north and 
limb-brightened emission is seen at south.
The medium image shows widespread emission distributed from
north to southeast, in addition to the southern bright shell.
In the hard energy band, 
several clumpy sources in the north region stand out
dramatically (Figure~\ref{fig:ximage}{\it c}). 
Hereafter, we refer to the clump centered
at a celestial coordinate (J2000) of 
\emph{R.A.}$=20^{\rm h}21^{\rm m}12^{\rm s}$,
\emph{decl.}$=40\degr 47\arcmin 55\arcsec$ as C1 and the other one at 
\emph{R.A.}$=20^{\rm h}20^{\rm m}00^{\rm s}$,
\emph{decl.}$=40\degr 40\arcmin 03\arcsec$ as C2.
Since another hard source C3 is located at the western edge of the FOV,
we do not report a spectral study of C3 in this paper.

The hard source C2 appears to be extended by a few arcminutes.
In order to evaluate quantitatively,
we compared the count rate profile centered on the clump C2
with the detector PSF in the 3--10~keV band, 
as plotted in Figure~\ref{fig:radial}.
The background level shown there is taken 
from the southern part of the remnant.
It is obvious that the hard X-ray emission extends up to 
4\arcmin --6\arcmin\ beyond the radial profile of a single point source.
In Figure~\ref{fig:radial}, we also plot the simulated profile of 
a circular source which has uniform surface brightness
with a radius of 4\arcmin\ centered on C2.
On the other hand, 
we obtain no sign of spatial extension of the source C1.

In order to compare the X-ray distribution with radio synchrotron emission,
the X-ray images on gray scale are overlaid with the 4850 MHz radio contours
in Figure~\ref{fig:ximage}.
The X-ray sources at the north coincide
with the radio-bright region and the southern X-ray shell with a 
fainter radio arc.
The X-ray morphology in general
bears close resemblance to the radio on large scales, 
but the X-ray emission in the proximity of the region R2
appears to anticorrelate with the radio emission.
The DR4, brightest radio region at the southeast, 
is quite dim in the X-ray band.
It is noticeable that there are enhancements
of radio emission possibly associated with the hard sources C1 and C2.

The second \citep{SecondEGRET}
and third \citep{ThirdEGRET} EGRET catalog report somewhat different 
positions for the unidentified $\gamma$-ray source inside $\gamma$~Cygni.
In Figure~\ref{fig:ximage}{\it c} we superimpose their 95\% confidence
error contours which are approximated as circles.
Whereas \citet{Brazier} reported a point-like X-ray source RX~J2020.2+4026 
by ROSAT close to the remnant center and within the EGRET error circle,
no point sources were detected there by our \emph{ASCA} observations.
The flux of RX~J2020.2+4026, estimated to be $\sim4\times10^{-14}~\flux$ 
assuming a photon index of 2, 
is indeed below the sensitivity of our observation.
In \S\ref{sec:nonthermal} we discuss a possible connection between 
the hard X-ray sources and the EGRET unidentified $\gamma$-ray emission.

\subsection{X-ray Spectra}\label{sec:xspec}
For the purpose of spectroscopic studies of the X-ray emissions
associated with the $\gamma$~Cygni SNR, 
contaminating emission that is overlapping the remnant 
must be taken into account adequately because 
the line of sight toward $\gamma$~Cygni passes along 
the Orion-Cygnus spiral arm.
Because of the lack of a blank field in our observations,
we analyzed four \emph{ASCA} archive data 
in neighboring fields (Field~1--4 in Table~\ref{tbl:cygnus})
to estimate the X-ray emission at
the $\gamma$~Cygni field unrelated to the SNR itself.
For each field, the GIS spectrum is obtained by integrated 
over the central detector region with a radius of 20\arcmin\ 
after eliminating resolved sources.
Each spectrum is subtracted by the high-latitude blank-sky spectrum
as a sum of the cosmic X-ray background and the instrumental background.
Figure~\ref{fig:cygnus} shows the resulting spectra 
in the 1.2--2.5~keV and 3.5--8~keV band,
after the correction for the integration sky area.
We find similar energy spectra from these neighborhood fields, 
except for Field~2 where the Galactic latitude is highest and 
the Galactic column density is lowest as compared with the other fields.
Once Field~2 is excluded, field-to-field variations of spectral data  
in the soft band are found to be insignificant.
The hard 3.5--8~keV spectrum of the southern field of $\gamma$~Cygni
is in agreement with those of Field~3 and 4 within their 
statistical uncertainties.
Consequently, we consider that Field~3 and 4 provide us 
a good approximation of the contaminating emission that should be
subtracted from the spectral data of $\gamma$~Cygni.
Since no bright sources are found in the \emph{ASCA} data of Field~4,
we have chosen Field~4 as ``background'' field.

We derive energy spectra of regions R1--R3 and clumps C1 and C2.
The clump spectra are extracted from 
the circular regions of a radius of 6\arcmin .
We exclude photons falling within 6\arcmin\ radius 
centered on C1 and C2 from the spectrum of the region R1.
Each accumulated on-source spectrum is subtracted by the 
``background'' Field~4 spectrum that is extracted from an 
identical detector region to each on-source data.
To improve statistics, spectra of two GIS detectors are always added.
Background-subtracted spectra of R1/R3 and C2
are shown in Figure~\ref{fig:xspec} and Figure~\ref{fig:xspec2},
respectively.
Several emission lines of Mg K ($\simeq 1.4$ keV)
and Si K ($\simeq 1.9$ keV) are evident in every spectra, indicative of
thin thermal plasma with a typical temperature of $\approx 1$ keV.
Remarkably, the spectrum of C2 exhibits
very flat continuum emission above 3~keV.
There are known errors in the calibration of the 
conversion of the photon energies to the pulse invariant channels
of the GIS detector below the xenon-L edge of 4.8 keV.
In the following spectral fitting, an artificial energy shift of $-50$ eV 
to each applied model is introduced to alleviate these errors
\citep{Buote}.

We have first attempted to fit the 0.7--8 keV spectrum of R3
by a thin thermal plasma model \citep{Mewe85,Liedahl90} 
in which collisional equilibrium ionization (CEI) is realized.
Photoelectric absorption along the line of sight is taken into 
account using the cross-sections from \citet{wabs}.
Elemental abundances are fixed to the solar values of \citet{Angr} 
throughout this paper unless otherwise mentioned.
The CEI plasma model cannot give an acceptable fit 
owing to a large discrepancy between the actual data and the 
model between Mg and Si K emission lines. 
Even when the abundances of alpha elements Ne, Mg, Si, S, and Fe
are individually allowed to vary ranging from 0.1 to 10 solar, 
the spectral data cannot be described properly.
In order to model emission line features,
we take account of the effects of non-equilibrium ionization \citep{Itoh}
by adopting a plasma emission code based on \citet{Masai}.
In an NEI plasma, degree of ionization and consequently 
line emissivities depend on 
the ionization timescale $n_e t$, where $n_e$ 
represents the electron density and $t$ the passage time after shocked.
The NEI plasma model yields an acceptable fit with 
a reduced $\chi^2 ({\rm d.o.f})=1.19(26)$.
The best-fit temperature and ionization timescale 
with $1\sigma$ errors are $kT_e = 0.76\err{0.10}{0.09}$~keV and 
$n_e t = 5.8\err{1.2}{1.4}\times10^{10}~{\rm cm}^{-3}{\rm s}$, 
respectively.

We found the R1 spectrum, as compared with R3, shows 
distinctive features, namely 
a strong emission at $\simeq 0.9$ keV and a hard continuum above 3~keV.
The former is consistent with the fact that R1 is 
very bright particularly in the 0.7--1~keV energy band;
the latter could be contamination by the hard sources C1 and C2.
Regarding the spectral fit of R1,
we include helium-like neon (\ion{Ne}{9}) K$\alpha$ line (0.923~keV),
in addition to the CEI plasma model that predominantly describes 
the 1--3~keV emission.
The CEI plasma plus \ion{Ne}{9} line model, however, 
cannot give a statistically acceptable fit, owing to 
residuals by the hard continuum.
Then, by adding power-law as a third component to describe the hard 
continuum, we obtain an acceptable fit for the R1 spectrum 
as summarized in Table~\ref{tbl:fit}.
The best-fit temperature is $kT_e = 0.56\err{0.03}{0.05}$~keV;
the photon index of the power-law component is $\Gamma =1.2~\err{1.1}{0.8}$.

The spectrum of the region R2 is fitted by a CEI
plasma model alone. The simple model yields an acceptable fit with 
the temperature of $kT_e = 0.53\pm0.07$~keV, which is
in complete agreement with the value obtained for the region R1.

The spectral data of the clump C2
indicate the presence of a hard continuum emission 
in the 3--8~keV energy band.
The hard emission is seen in the spectrum of C1 as well.
As the case of R1, we employ a three-component model comprised of 
the \ion{Ne}{9} line (0.92~keV), CEI thermal plasma, and power-law spectrum.
We find that a good fit cannot be obtained if
we omit the power-law component.
We fitted the three-component model to the 0.7--8~keV spectral data 
of the clumps C1 and C2, by freezing 
the plasma temperature to the best-fit value
of the region R1 and attenuating
all components with a common absorption column of 
$N_{\rm H}= 0.84\times10^{22}~{\rm cm}^{-2}$ that is also obtained for R1.
The power-law components are found to be very flat;
the best-fit photon indices are
$\Gamma = 1.5\pm{0.5}$ and $0.8\pm{0.4}$ 
for C1 and C2, respectively.
Instead of adding a power-law, we also attempted to add a 
thermal bremsstrahlung component to model the high-energy part 
of the C2 spectrum.
The 90\% lower-limit on the temperature of 
the thermal bremsstrahlung emission is set to be $kT_e = 5.8$ keV.

\section{Discussion}\label{sec:discuss}

\subsection{Estimation of SNR Distance and Age}\label{subsec:sedov}
\citet{Higgs77} derived the distance to $\gamma$~Cygni as $1.8\pm0.5$~kpc
based on the $\Sigma$--$D$ relation which 
is a statistical property of the radio brightness of SNRs.
\citet{Land80} estimated as $1.5\pm0.5$~kpc and pointed out 
the progenitor of the $\gamma$~Cygni remnant was possibly a member of 
the Cyg~OB9 association at $1.2\pm0.3$~kpc.
The absorbing column density provides additional information about 
the distance.
\citet{Maeda} reported Wolf-Rayet binary \VCYG\ located close to 
the $\gamma$~Cygni sky field is attenuated by 
the interstellar column density of
$N_{\rm H}=(1.1\pm0.2)\times10^{22}~{\rm cm}^{-2}$, 
similar to the $\gamma$~Cygni remnant.
Thus, we expect the distance to $\gamma$~Cygni is probably
close to the distance to \VCYG , 1.7 kpc.
In view of these arguments, we take the distance $D=1.5$~kpc as most 
probable value.

The X-ray distribution of the region R3 shows clear arc-like 
morphology along the outer boundary of the
shell structure observed at radio frequencies.
The X-ray spectrum was modeled by a single temperature 
thermal emission with the solar abundance.
These features are indicative of a thin thermal plasma 
in the immediate postshock region
of the primary blast wave propagating through the interstellar matter.

Assuming the equipartition between the shocked electrons and ions
and Rankine-Hugoniot jump conditions for a strong shock,
the electron temperature is related to the shock velocity $v_s$ as
$kT_e =(3/16) \mu m_p v_s^2$, where $\mu =0.6$ is a mean mass per 
particle in units of the proton mass $m_p$.
The best-fit value of $kT_e = 0.76\err{0.10}{0.09}$~keV 
corresponds to the shock velocity $v_s = 800\err{50}{60}$ km/s.

Provided that the remnant is in the Sedov adiabatic expansion phase,
the age of $\gamma$~Cygni can be estimated to be
$\tau_{\rm age} = (2/5) R/v_s$, where
$R$ is the radius of the supernova shock front.
Given the angular radius of $\theta \simeq 30\arcmin$,
we have the physical radius $R= D\theta \simeq 13.5 D_{1.5}$~pc,
where $D_{1.5}$ is the distance in units of $1.5$~kpc.
Thus, we obtain an age estimate of $\tau_{\rm age} \simeq 6600 D_{1.5}$ yr.

\subsection{Signatures of Shock--Cloud Interaction}\label{subsec:cloud}
On the basis of the \ion{H}{1} line emission and absorption 
toward the $\gamma$~Cygni SNR,
\citet{Land80} suggested that the supernova explosion took place in 
the slab of an interstellar cloud oriented north to southeast.
The soft (1--3~keV) X-ray emission ``belt'' from north to southeast
appears to agree fairly well with 
the spatial distribution of the \ion{H}{1} line features.
A possible scenario explaining this coincidence is that 
the X-ray emission belt is due to 
thermal evaporation of clouds \citep{WhiteLong} as a consequence of 
the collision between the supernova blast wave and the dense clouds.
The supernova shock may expand into a cavity 
produced by progenitor stellar winds, and then encounter the cavity wall
comprised of the clouds \citep{Chevalier99}.

We found the possible evidence of the strong emission line of 
helium-like Ne from the R1 spectrum.
The X-ray map in the 0.7--1~keV band suggests clumpy nature of 
the \ion{Ne}{9} line emission.
The upper limit for the hydrogen-like Ne line intensity 
is found to be 6.7\% of the helium-like intensity.
This gives an upper limit of 0.24~keV for the temperature of plasma
emitting the \ion{Ne}{9} line if we assume ionization equilibrium state.
The low-temperature and clumpy appearance are indicative 
of the shock--cloud interactions at the northern part of the remnant.

The hard sources C1 and C2 reside in the northern region
where the shock--cloud interaction is likely to occur.
Therefore, we identify the hard sources
with shocked dense cloudlets.
This speculation may be encouraged by the presence of 
high-velocity \ion{H}{1} cloudlets 
with a density of 10--100~cm$^{-3}$ and 
an apparent radius of 4\arcmin --6\arcmin\ 
in the southeast of the remnant \citep{Land80}.
Let us calculate the density of plasma that is responsible
for the derived Ne flux 
$I_{\rm Ne}\sim 6(0.8)\times10^{-3}$~photons~cm$^{-2}$s$^{-1}$
of the clump C1(C2) with reasonable parameters.
Given a temperature of 0.1~keV and a spherical volume of an angular
radius of 2\arcmin\ (assuming distance $D=1.5$ kpc),
we have the density $n \approx 60(8)$~cm$^{-3}$ by using 
the Ne line emissivity of \citet{Mewe85}, in agreement with 
the typical density of \ion{H}{1} cloudlets.

\subsection{On the Origin of Hard X-ray Emission}\label{sec:nonthermal}
An important finding of our observations is 
the clumpy hard X-ray emissions from the regions C1 and C2
(hereafter, HXC -- the hard X-ray clump). 
We found that C2 has an extended structure with an effective radius of 
about $4\arcmin$.
The HXC components are extremely hard and 
characterized by a power-law with $\Gamma \sim$~0.8--1.5.
As we argued above,
it is likely the HXC originates in shocked dense cloudlets
with a gas density between 10 and 100 $\rm cm^{-3}$.

The thermal-bremsstrahlung interpretation of the  
high-energy part of the C2 spectrum requires 
very high temperature ($kT_e >5.8$~keV).  Since the 
shocks of an evolved SNR are not sufficiently energetic  
to heat the gas to such temperatures, the nonthermal 
origin of the hard X-ray emission seems a more favorable option. 
The synchrotron radiation, the inverse-Compton scattering, and 
the nonthermal bremsstrahlung of shock-accelerated electrons 
are three natural production mechanisms of hard X-rays.
The overall flux of the HXC in the 2--10~keV interval 
is estimated as $F_{\rm X} \simeq 4.5\times 10^{-12}\flux$, which corresponds 
to the source luminosity  
$L_{\rm X} \simeq 1.2\times 10^{33}D_{1.5}^2~\lumi$.

The nonthermal luminosity of this SNR peaks at high energy $\gamma$-rays,
if the association with the EGRET source is real.
The $\gamma$-ray spectrum measured by EGRET is shown 
in Figure~\ref{fig:nufnu}, together with the HXC spectrum.
The observed photon flux of the EGRET source
is translated into a luminosity of 
$L_{\gamma} \simeq 1.4\times10^{35}D_{1.5}^2~\lumi$ (100~MeV--2~GeV)
for the photon index $2.1$.
The reported positions of the EGRET $\gamma$-ray 
source 2EG~J2020+4026 and/or 3EG~J2020+4017 do not coincide 
exactly with the HXC.
However the large systematic errors in the EGRET position, which
strongly depend on the chosen diffuse  $\gamma$-ray emission  
model \citep{Hunter97}, do not allow certain conclusions
concerning the location of the $\gamma$-ray production region. 
Therefore it is interesting to test whether the HXC
can be a counterpart of the unidentified GeV $\gamma$-rays 
source on theoretical ground. 

\subsubsection{Coulomb Cooling and Nonthermal Bremsstrahlung}
\label{subsec:accel}
The emission region of the HXC is supposed to be the 
dense cloudlet engulfed by the supernova blast wave,
and therefore the density is probably high ($n \sim$~10--100~cm$^{-3}$).
Below we assume that accelerated electrons are continuously 
injected in the HXC, by the accelerator(s) being located either outside
or inside the HXC.
Possible internal accelerator could be the secondary shocks
generated by the interaction between 
the primary blast wave and the dense cloudlets.
Indeed, \citet{Bykov} considered electron acceleration 
at a slow shock (the order of 100~km/s) transmitted inside a dense cloudlet.
A tail shock could be formed behind the cloudlets \citep{Jones93} as well,
although an efficient particle acceleration at the tail shock is not obvious. 
On the other hand, the external accelerator should be 
the primary shock of the SNR.

The energy spectrum of accelerated electrons depends 
on the acceleration mechanism,
the discussion of which is beyond the scope of this paper. 
For simplicity we assume the electron production rate 
in momentum space
is described by a single power-law function $Q(p) \propto p^{-\kappa}$.
Such a function implies power-law distributions of electrons 
also in the kinetic energy space, although  
in the non-relativistic and ultrarelativistic regimes 
the power-law indices are different,
namely  $Q(E) \propto E^{-(s+1)/2}$ at $E \ll m_e c^2$ while
$Q(E) \propto E^{-s}$ at $E \gg m_e c^2$, where $s=\kappa+2$.
In the case of diffusive shock acceleration at the shock front,
according to the standard model \citep{Bell78a, BlandfordEichler87}, 
the spectral index is determined by the compression ratio of the shock as 
$s=(r+2)/(r-1)$. For strong shocks with $r \sim 4$
the spectral index is close to 2.    

For simplicity we suppose that the electrons are 
effectively trapped in HXC; thus the present-day electron 
spectrum is formed by accumulation of 
freshly accelerated (or arriving from external accelerators) 
electrons during the entire history of the HXC.
If the energy-losses can be neglected, 
the electron spectrum becomes $N(E) = \tau_{\rm age} Q(E)$, 
where the age of the accelerator is approximated by
the age of $\gamma$~Cygni, i.e. $\tau_{\rm age} \simeq 7000$~yr. 
It is valid, however, only in the intermediate energy
region, typically between 300 MeV to 10 GeV, 
where the bremsstrahlung dominates the radiative losses.
Indeed, the characteristic lifetime of electrons against 
the bremsstrahlung losses, 
$\tau_{\rm br}=E/(-dE/dt)_{\rm br} 
\simeq 4.3 \times 10^7 (n/10 \ \rm cm^{-3})^{-1}\rm ~yr$ is 
considerably larger than the age of $\gamma$~Cygni as long as the gas density 
does not exceed $6\times 10^4~\rm cm^{-3}$. 

The high density and presumably strong magnetic field in the HXC
make, however, the cooling processes  crucial  in the formation 
of the electron spectrum at low and very high energies. 
The relevant cooling processes are the Coulomb (or, in the neutral gas,
ionization) losses for low-energy electrons and the synchrotron 
losses for high-energy electrons.    
The inverse-Compton losses could be omitted if the  
magnetic field in HXC exceeds $10~\mu \rm G$. Also, it should be 
noticed that even in case of significant bremsstrahlung losses 
($\tau_{\rm br} \leq \tau_{\rm age}$),
the process does not modify the electron spectrum. 

Below several hundreds MeV, the energy losses of electrons are 
dominated (independent of the gas density) 
by the Coulomb losses \citep{Hayakawa69}.
The cooling time of electrons due to the Coulomb interactions 
is given by \citep{Rephaeli}
\begin{eqnarray} 
\tau_{\rm cou} &=& 
\frac{E}{(3/2)n_e c \thomson (m_e c^2) \beta^{-1} \ln\Lambda}\\
        &=& 4.3 \times 10^3
        ~\beta \left( \frac{n}{10~{\rm cm}^{-3}} \right)^{-1}
        \left( \frac{E}{{\rm MeV}} \right) ~{\rm yr},
\end{eqnarray}
where $\thomson$ is the Thomson cross-section,
$c$ is the velocity of light, $E$ and $\beta$ are the kinetic energy and 
the velocity of nonthermal electrons in units of $c$, 
$m_e$ is the electron mass, and 
$\ln\Lambda$ is the Coulomb logarithm which is set to be 40. 
Equating $\tau_{\rm cou}$ and $\tau_{\rm age}$ gives the break energy 
$E_{\rm cou}$, ``Coulomb break'', 
below which the electron spectrum is flattened:
\begin{equation} \label{eq:E_cou}
E_{\rm cou} \sim 1.6~
        \left( \frac{n}{10~{\rm cm}^{-3}} \right)
        \left( \frac{\tau_{\rm age}}{7000~{\rm yr}} \right)
        ~{\rm MeV},
\end{equation}
where we use the approximation of $\beta \sim 1$.
At energies below $E_{\rm cou}$ 
the Coulomb losses significantly modify the 
acceleration spectrum, $N(E) = \tau_{\rm cou}(E)Q(E)$;
namely, at relativistic energies, $N(E) \propto E^{-s+1}$.

The differential energy spectrum of the bremsstrahlung emission from
these electrons is calculated as \citep{Blumenthal}
\begin{equation}
I(\varepsilon) \simeq
\int dE N(E)c\beta
\left( n_{\rm H}\frac{d\sigma_{e\rm H}}{d\varepsilon}
+ n_{\rm He}\frac{d\sigma_{e\rm He}}{d\varepsilon}
+n_{e}\frac{d\sigma_{ee}}{d\varepsilon} \right) ,
\label{eq:bremss}
\end{equation}
where $n_{\rm H}$, $n_{\rm He}$, and $n_{e}$ are 
hydrogen, helium, and electron number densities, respectively, 
and $d\sigma/d\varepsilon$ is the differential cross-section
for emitting a bremsstrahlung photon in the energy interval $\varepsilon$ to 
$\varepsilon+d\varepsilon$.
We adopt ratios $n_{\rm He}/n_{\rm H}=0.1$ and $n_{e}/n_{\rm H}=1.2$.
In the ultrarelativistic regime the 
electron-electron bremsstrahlung becomes
comparable to the electron-proton bremsstrahlung.
Since the bremsstrahlung cross-section is in inverse proportion to
the emitted photon energy,
$d\sigma/d\varepsilon \propto \varepsilon^{-1}$, and only 
slightly (logarithmically) depends
on the electron energy in the relativistic regime, 
the bremsstrahlung photon spectrum
almost repeats the power-law spectrum of parent electron
distribution of $N(E)=kE^{-\alpha}$,
i.e. $I(\varepsilon) \propto  \varepsilon^{-\alpha}$ if $\alpha \geq 1$.  
Then, we should expect a broken power-law spectrum with  
$I(\varepsilon) \propto  \varepsilon^{-s}$ at 
$\varepsilon \geq  E_{\rm cou}$, 
and $I(\varepsilon) \propto  \varepsilon^{-s+1}$ 
at $\varepsilon <  E_{\rm cou}$ in the relativistic domain.    
The energy spectrum of bremsstrahlung photons 
is essentially determined by the gas density and the age.
Below the ``Coulomb break'',
for any reasonable acceleration index of $s \leq 2.5$,
we see a hard differential spectrum with a photon index less than 1.5.
It should be stressed that,
for the electron spectrum harder than $E^{-1}$
($E^{-1/2}$ in the non-relativistic case), 
the bremsstrahlung photons obey a standard $\varepsilon^{-1}$ type spectrum.
In the \emph{ASCA} energy band, we should always expect the standard spectrum.
 
\subsubsection{Synchrotron Radiation} \label{sebsec:sync}
The synchrotron cooling time of electrons in the magnetic field 
$B$ is given by 
\begin{eqnarray} \label{eq:tau_sync}
\tau_{\rm syn} &=& \frac{\gamma m_e c^2}{(4/3) c \thomson U_B \gamma^2} \\
        &=& 1.2 \times 10^{3} 
        \left( \frac{B}{10^{-5}\rm G} \right)^{-2}
        \left( \frac{E}{10^{14}\rm eV} \right)^{-1} 
        ~{\rm yr}, 
\end{eqnarray}
where $\gamma$ is the electron Lorentz factor and $U_B=B^2/(8\pi)$ is 
the energy density of the magnetic field.
Equating $\tau_{\rm syn}$ and $\tau_{\rm age}$ gives the break energy 
$E_{\rm syn}$ above which the electron spectrum is steepened:
\begin{equation} \label{eq:sync_break}
E_{\rm syn} \sim 18~
        \left( \frac{B}{10^{-5}\rm G} \right) ^{-2}
        \left( \frac{\tau_{\rm age}}{7000~{\rm yr}} \right) ^{-1}
        ~{\rm TeV}.
\end{equation}

Thus if the maximum energy of electrons exceeds 
$E_{\rm syn}$, we should expect a break in the electron spectrum. 
In particular, in the case of power-law acceleration spectrum,  
the loss-steepened spectrum has a power-law form 
$N(E) =\tau_{\rm syn}(E)Q(E) \propto E^{-s-1}$.

The average synchrotron photon energy radiated by an electron 
with energy of $E_{14}$ (in $10^{14}$~eV) 
in the field $B_{-5}$ (in $10^{-5}$~G) is
$\epsilon \sim 1.6 B_{-5} E_{14}^2$~keV.
Then, the break energy $E_{\rm syn}$ 
at which the synchrotron lifetime is balanced with the SNR age
is translated into the synchrotron photon energy of 
$\epsilon_{\rm syn} \sim 50~ (B/10^{-5}{\rm G})^{-3} (\tau_{\rm age}
/7000~{\rm yr})^{-2}~{\rm eV}$,
above which noticeable spectral steepening should appear. Therefore   
the observed flat X-ray spectrum up to 10~keV 
would require unrealistically small magnetic field in the HXC region,  
$B \leq 2~\mu \rm G$.
  
Furthermore, it is easy to show that in the case of 
acceleration of electrons inside HXC, independent of the strength of 
the magnetic field, the cut-off energy in the 
synchrotron spectrum cannot exceed a few eV.
Indeed, if the  maximum energy of accelerating electrons 
is limited by the synchrotron losses, 
the balance between  the acceleration and cooling timescales
determines the  maximum electron energy. 
The timescale of the first-order Fermi acceleration  
in the Bohm diffusion limit is given by
\begin{eqnarray}
\tau_{\rm acc} &\sim& \frac{r_L c}{v_s^2} \\ 
        &=& 3.1\times 10^{5} 
        \left( \frac{B}{10^{-5}\rm G} \right)^{-1}
        \left( \frac{v_s}{100~\rm km/s} \right)^{-2}
        \left( \frac{E}{10^{14}\rm eV} \right) 
        ~{\rm yr}, 
\end{eqnarray}
where $r_L=m_e \gamma / (eB)$ is the Larmor radius of electrons 
and $v_s$ is the shock velocity. 
Equating $\tau_{\rm syn}$ in equation~(\ref{eq:tau_sync})
and $\tau_{\rm acc}$ gives the maximum energy 
$E_m \sim 6.2~ (B/10^{-5}{\rm G})^{-1/2} (v_s/100~{\rm km/s})$~TeV,
and its characteristic photon energy is 
$\epsilon_{m} \sim 3.8~(v_s/100~{\rm km/s})^2$~eV that is
independent of the magnetic field strength.
Hence, regardless of the magnetic field strength, 
the first-order Fermi acceleration mechanism 
cannot explain the extremely flat X-ray spectrum
by synchrotron radiation, 
unless the shock speed exceeds 
$5000 \ \rm km/s$. The latter condition obviously cannot be
satisfied in $\gamma$~Cygni.

\subsubsection{Inverse-Compton Emission}
The hard X-rays could be produced also by inverse-Compton (IC) 
scattering of seed photons off relativistic electrons.
In the $\gamma$~Cygni SNR, 
the cosmic microwave background (CMB) radiation dominates over 
other ambient photon fields like the diffuse Galactic 
infrared/optical radiation  and the infrared emission from shock-heated dust
\citep{Gaisser98}.

Relativistic electrons which boost the CMB photons up to X-rays 
also emit synchrotron radiation in the radio band.
Characteristic energies of synchrotron photon $\epsilon_{\rm syn}$
and IC photon $\epsilon_{\rm ic}$ (in keV) 
produced by the same electrons are related as \citep{Aharonian97}
\begin{equation}\label{eq:AAK}
\epsilon_{\rm syn}\simeq 
0.7\times10^{-7} \epsilon_{\rm ic} B_{-5}~{\rm eV}
\end{equation}
for the CMB seed field.
The ratio  of the spectral power 
$f(\varepsilon)\equiv \varepsilon^2 \times I(\varepsilon)$
of the IC X-ray emission to the 
synchrotron emission at the relevant radio frequency is 
\begin{equation}\label{eq:ic2syn}
\frac{f_{\rm ic}(\epsilon_{\rm ic})}{f_{\rm syn}(\epsilon_{\rm syn})}
=\frac{U_{\rm CMB}}{U_{B}},
\end{equation}
where $U_{\rm CMB}=0.25~{\rm eV}~{\rm cm}^{-3}$ is the energy density of 
the CMB radiation.
The spectral power of the observed radio emission from the whole remnant,
in the frequency range from 408 MHz to 4.8 GHz 
\citep{Wendker91,Higgs77} is well described as 
\begin{equation}\label{eq:radio}
f_{\rm radio}(\varepsilon) = 5\times10^{-12} 
\left( \frac{\varepsilon}{10^{-5}~{\rm eV}} \right)^{0.5} \flux .
\end{equation}
Since almost 60\% of the radio flux comes from the southeast 
of $\gamma$~Cygni, from equation~(\ref{eq:AAK}) and (\ref{eq:ic2syn}) 
we find that the IC spectral power 
cannot exceed  $1.7\times 10^{-14}\epsilon_{\rm ic}^{0.5}
B_{-5}^{-1.5}~\flux$.
This upper-limit is less than $1 \%$ 
of the flux of the hard X-ray emission detected.  
Thus, the IC radiation hardly  can explain the flux of HXC, unless 
the magnetic field is extremely weak ($< 1 \ \mu \rm G$).

\subsubsection{Nonthermal Bremsstrahlung X/$\gamma$-radiation}
The nonthermal bremsstrahlung from the accelerated electrons
is a natural source of the HXC flux,
because the shocked dense cloudlets act as an effective 
target for energetic electrons. Because of Coulomb interactions 
the high density gas leads
to significant hardening of low-energy electrons below the ``Coulomb break'',
giving rise to the standard $\varepsilon^{-1}$ type
bremsstrahlung spectrum at the X-ray band, 
which perfectly agrees with the \emph{ASCA} data.
The bremsstrahlung spectrum above the ``Coulomb break'' essentially repeats 
the acceleration spectrum of electrons.
For the given spectral index of accelerated electrons,  
the ratio of the X- and $\gamma$-ray fluxes depends only on 
the ``Coulomb break'' energy
which in its turn depends on the gas density and the 
age of the accelerator.   

In Figure~\ref{fig:nufnu} we show the results of numerical calculations 
for two sets of parameters which describe the gas 
density and the acceleration spectrum of electrons, by assuming
the electron bremsstrahlung is responsible for both  
the \emph{ASCA} hard X-ray and the EGRET $\gamma$-ray fluxes. 
For the electron spectrum with the acceleration index $s=2.1$,
the best fit is achieved for a gas density of $n = 34 \ \rm cm^{-3}$.
A steeper acceleration spectrum with $s=2.3$ requires larger gas 
density, $n = 130 \ \rm cm^{-3}$. 
Note that the adopted acceleration spectra are consistent with 
the reported radio spectral index $\alpha = 0.5\pm 0.15$.
We also suppose an exponential cutoff in the electron spectrum at 10 TeV. 

If the electron distribution with $s=2.3(2.1)$ extends beyond GeV energies,
for the magnetic field $10^{-5}$~G the calculated 
radio flux density
amounts to about 10\%(60\%) of the measured radio flux density
integrated over the whole remnant.
Furthermore, if the electron distribution extends beyond TeV,
we found the bremsstrahlung spectrum with $s=2.1$ exceeds  
the Whipple upper-limit, whereas the spectral index of $s=2.3$ is still 
in agreement with the Whipple data. 
Meanwhile both combinations of model parameters satisfactorily 
fit the spectral shape and the absolute flux of hard X-rays. 
The ignorance of energy losses of electrons 
would lead to significantly steeper X-ray spectra, and would 
also result in overproduction of absolute X-ray fluxes (dotted curves in 
Figure~\ref{fig:nufnu}).
Note that the main contribution to X-rays comes from 
relatively high energy electrons with energies close to 10~MeV
for the electron index $s=2.3$.

Because of poor angular resolution, the EGRET measurements 
do not provide a clear information about the 
site(s) of production of high energy $\gamma$-rays. Nevertheless, 
it is likely that only a part (perhaps, even only a small part)  
of the reported high energy $\gamma$-ray fluxes
originates in the HXC region.    
The $\gamma$-ray fluxes could be suppressed 
by assuming lower gas  densities. 
Indeed, such an assumption would lead to the shift of the ``Coulomb break'' 
energy in the electron spectrum 
to lower energies, and the predicted high energy $\gamma$-ray spectra  
would appear significantly below the reported EGRET fluxes
(a solid curve in Figure~\ref{fig:nufnu}).

A more likely candidate for production of the bulk of high energy 
$\gamma$-rays is the region called DR4 from which most of the radio 
emission emerges.
A massive cloud with a density of $\sim 300$~cm$^{-3}$ 
occupying $\sim 5$\% of the SNR volume has been suggested to exist 
in the vicinity of DR4 to explain the $\gamma$-ray flux \citep{Pollock}.
Actually the EGRET error circle reported is somewhat away from the HXC
but closer to the DR4.
A density of $\sim 300$~cm$^{-3}$ is higher than the upper limit 
density of the HXC. Such high gas density implies 
a high (about 50 MeV) ``Coulomb break'' energy in the electron spectrum,
and therefore considerable suppression of X-ray flux.
This could naturally explain the lack of noticeable hard X-ray fluxes from 
the DR4 region which is bright in radio and possibly $\gamma$-rays.

Finally, we briefly discuss the power consumption due to the rapid 
Coulomb losses.
For the gas density of about $100 \ \rm cm^{-3}$ in the HXC,
the X-ray flux is produced predominantly by electrons with energies 
of about 10~MeV.
The X-ray flux is roughly proportional to the product of the gas density 
and the number of relativistic electrons,
because the relativistic bremsstrahlung cross-section 
depends only logarithmically on the electron energy.  
On the other hand, the Coulomb energy loss rate of relativistic electrons,
$dE/dt \propto E/\tau_{\rm cou}$, 
is proportional to the gas density and almost independent of the 
electron energy.
Therefore the energy loss rate of the bulk of the X-ray emitting electrons 
can be uniquely determined by the X-ray luminosity.
The measured X-ray luminosity,
$L_{\rm X} \simeq 1.2\times 10^{33}D_{1.5}^2~\lumi$,
can be converted to the energy loss rate of 
$L_e \sim 5\times 10^{37}D_{1.5}^2~\lumi$.
The energy released in relativistic electrons is roughly estimated as
$W_e \sim \tau_{\rm age} L_e \sim  10^{49}D_{1.5}^2$~ergs.
This enormous energy deposition due to Coulomb collisions
would heat the emission region of the HXC.
Subsequently the heat would be radiated away in the far-infrared band
by molecular line emission, if the gas is comprised of molecules.
The observed infrared luminosity of $\gamma$~Cygni
\citep{Saken} is comparable to the energy loss rate estimated above.
On the other hand, if the emission region of the HXC are predominantly 
shock-ionized plasmas, the deposited energies by the accelerated 
electrons would heat up the plasmas.

\section{Summary and Conclusion}
The X-ray emissions emerging from the $\gamma$~Cygni SNR 
are found to be complex.
Spatial and spectral studies of the X-ray data
reveal the presence of four components:
(1) limb-brightened thermal emission with a temperature of $\simeq 0.8$~keV;
(2) widespread emission with a temperature of $\simeq 0.6$~keV
aligned from north to southeast bounded by the radio-bright regions;
(3) strong emission lines from \ion{Ne}{9} ions in the vicinity of
the northern bright-radio region;
(4) clumpy hard emissions, which is best described by unusually hard 
power-law spectral distribution with a photon index  $\Gamma \simeq $ 0.8--1.5.

The limb-brightened component is considered to be thermal plasma 
emission from the immediate postshock region
of the SNR blast wave that is propagating through the interstellar medium.
The temperature gives the age estimate of $\tau_{\rm age}\sim 6600$~yr
based on the Sedov evolution of the remnant.
The soft X-ray emissions from north to southeast 
are likely to be caused by the interaction between the supernova 
shock and the cavity wall comprised of ambient clouds.
In particular, the intense neon line emission would be attributable to 
the low-temperature plasma generated by the 
shock--cloud interaction.

The extremely hard X-ray emission of the HXC is naturally explained 
by the nonthermal bremsstrahlung from the loss-flattened electron distribution.
By assuming that the HXC is a counterpart
of the EGRET unidentified $\gamma$-ray source 3EG~J2020+4017,
we estimate the density of the emission region to be 
$\sim 130$~cm$^{-3}$, for the electron index 2.3.
The bremsstrahlung interpretation requires 
very large Coulomb energy loss rate of $\sim 5\times 10^{37}~\lumi$
and consequently the total amount of the energy loss of 
$\sim 10^{49}$~ergs,
regardless of the gas density and the energy of
electrons responsible for the hard X-radiation.
Given the limited energy budget of $\gamma$ Cygni,
it is more comfortable to attribute this huge energy release 
to the acceleration at the HXC region, 
rather than to assume that the electrons are supplied by 
external accelerators. 
Otherwise, the energy requirement would be increased
by an order-of-magnitude or more, unless we assume very specific 
arrangement of external accelerators around the HXC region.      
 
If the HXC region contributes only a fraction of the EGRET 
high energy $\gamma$-ray flux,
the gas density estimate of 130~cm$^{-3}$ 
should be considered as an upper limit.
All accelerated electrons are presumed to be trapped 
inside the HXC volume.
However, it may be possible that high-energy ($\geq 100$ MeV) 
electrons effectively escape from the HXC, and later interact with 
very dense clouds resulting in high energy 
$\gamma$-rays and synchrotron radio emission, but without noticeable
X-radiation. 
The spatial structures of the HXC will be comprehensively 
studied by our forthcoming \emph{Chandra} observation.
The origin of the $\gamma$-ray emission from $\gamma$~Cygni
will eventually be tested by the future  
\emph{Gamma-Ray Large Area Space Telescope} (\emph{GLAST}) mission.

\acknowledgments
We thank Dr. L. Drury for his critical comment concerning the 
interpretation of the site of electron acceleration.
Y.U. thanks Dr. M. Ishida for stimulating discussions and his 
advice on the \emph{ASCA} data analysis, and also wish to thank
all the members of the \emph{ASCA} team.
The work of Y.U. was supported in part by the Research Fellowships of
the Japan Society for the Promotion of Science for Young Scientists.

\clearpage


\figcaption[ximage.eps]{({\it a})--({\it c}):
\emph{ASCA} GIS X-ray images (\emph{gray scale})
of the $\gamma$~Cygni supernova remnant 
in ({\it a}) 0.7--1~keV, ({\it b}) 1--3~keV, and ({\it c}) 4--7~keV 
energy bands.
Two GIS detectors (GIS2, GIS3) are summed.
The brightness level indicated at the top of the images is 
in units of the surface brightness of 
the cosmic X-ray background of the relevant energy interval.
Radio contour map (NRAO 4.85~GHz: 7\arcmin\ FWHM) 
is superposed on the \emph{ASCA} images.
A central part of strong emission 
around an \ion{H}{2} region called $\gamma$~Cygni nebula 
at southeast is blanked from the radio map. 
Also shown are the EGRET 95\% confidence error circles 
(\emph{dotted circle}: 2EG~J2020+4026, \emph{thick circle}: 3EG~J2020+4017).
\label{fig:ximage}}

\figcaption[radial.eps]{The GIS radial count rate profile
of the clump C2 in the 3--10~keV energy band.
A histogram represents the simulated radial profile of a point source.
Estimated background level is taken from the southern part of the remnant.
\label{fig:radial}}

\figcaption[diffbgd.eps]{A comparison between the energy spectra 
integrated over the GIS field-of-view
in the 1.2--2.5~keV energy band (\emph{top}) 
and the 3.5--8~keV band (\emph{bottom});
the south $\gamma$~Cygni (\emph{filled circle}), Field~1 (\emph{open circle}),
Field~2 (\emph{rectangle}), Field~3 (\emph{diamond}), 
and Field~4 (\emph{triangle}).
\label{fig:cygnus}}

\figcaption[r13.eps]{The GIS energy spectra 
extracted from the regions R1 (\emph{filled circle}) 
and R3 (\emph{open circle}), with linear scale.
The R3 spectrum is multiplied by a factor of 0.3 for display purpose only.
The curves show best-fit models, 
folded through the response function of the instrument.
The bottom panels plot the residuals of data 
compared with the thermal emission models.
\label{fig:xspec}}

\figcaption[c2.eps]{The GIS energy spectra 
extracted from the clump C2 (\emph{filled circle}),
where the background data are taken from the Field~4.
The curve shows only the thermal component of the best-fit model,
folded through the response function of the instrument.
The bottom panel plots the residuals of data 
compared with the thermal component.
Also shown is the high-energy part of the C2 spectrum 
whose background data are taken from the sourthern portion of 
$\gamma$~Cygni (\emph{open circle}); 
the high-energy continuum is hardly affected by the
choise of the background data.
\label{fig:xspec2}}

\figcaption[broadband.eps]{Broadband spectral power of 
the $\gamma$~Cygni SNR.
The range of the power-law fit of the hard X-ray component    
is shown together with $\gamma$-ray data ($\geq 100$~MeV)
of 2EG~J2020+4026 taken from \citet{Esposito}
and the Whipple TeV upper limit from \citet{Buckley}.
The bremsstrahlung photon spectra from the loss-flattened electron
distribution are calculated for
the electron index $s=2.1$ and the gas density $n=34\ \rm cm^{-3}$
(\emph{long-dashed line}),
$s=2.3$ and $n=130\ \rm cm^{-3}$ (\emph{dashed line}), and
$s=2.3$ and $n=10\ \rm cm^{-3}$ (\emph{solid line}).
The \emph{dotted lines} show the bremsstrahlung 
spectra corresponding  
to the acceleration  spectra of electrons, i.e. 
ignoring the Coulomb losses of electrons.  
\label{fig:nufnu}}



\clearpage

\begin{deluxetable}{ccccc}
\tabletypesize{\small}
\tablecolumns{4} 
\tablecaption{Summary of the \emph{ASCA} archive fields
near the $\gamma$~Cygni SNR\label{tbl:cygnus}}
\tablewidth{0pt}
\tablehead{
\colhead{Field} & \colhead{Target} & \colhead{Coordinate} &
\colhead{Distance\tablenotemark{a}} \\
\colhead{}      &  \colhead{}      & \colhead{$(l,b)$} &
\colhead{(deg)} \\
}
\startdata
$\gamma$~Cygni \tablenotemark{b}
& 2EG~J2020+4026 2 & ( $77\fdg92,~2\fdg22$ ) & 0.0  \\
Field 1 & \VCYG & ( $76\fdg66,~1\fdg43$ ) & 1.5     \\
Field 2 & \NGC  & ( $75\fdg55,~2\fdg42$ ) & 2.4     \\
Field 3 & \GRO  & ( $75\fdg45,~0\fdg61$ ) & 2.9     \\
Field 4 & \GEV  & ( $81\fdg22,~1\fdg02$ ) & 3.5     \\
\enddata
\tablenotetext{a}{Angular distance from $(l,b)=(77\fdg92,~2\fdg22)$.}
\tablenotetext{b}{This pointing covers the southern part 
of the $\gamma$~Cygni SNR.}
\end{deluxetable}

\clearpage
\begin{deluxetable}{lccccc}
\tablecolumns{6} 
\tabletypesize{\small}
\tablecaption{Results of Spectral Fits to the \emph{ASCA} 
data of $\gamma$~Cygni
\label{tbl:fit}}
\tablewidth{0pt}
\tablehead{
\colhead{Parameter} & \colhead{R3} & \colhead{R2} &
\colhead{R1}& \colhead{C2} & \colhead{C1}\\
}
\startdata

Power Law: &  & & &\\
Photon Index $\Gamma$ &
\nodata &\nodata & 
1.2~$\err{1.1}{0.8}$ & 0.8~$\pm0.4$ & 1.5~$\pm0.5$ \\
$F_{2-10~{\rm keV}}$ \tablenotemark{a}  &
\nodata &\nodata & 
1.8~$\err{0.6}{0.5}$ & 0.98~$\err{0.21}{0.20}$ & 1.7~$\err{0.5}{0.4}$ \\
\\[0mm]

Thermal Plasma: & & & & & \\
$kT$~(keV) &
0.76~$\err{0.10}{0.09}$ & 0.53~$\pm0.07$ & 0.56~$\err{0.03}{0.05}$&
0.56~$\fixed$ & 0.56~$\fixed$ \\
EM~($10^{12}$cm$^{-5}$) \tablenotemark{b} &
3.5~$\err{1.0}{0.7}$ & 3.0~$\err{1.5}{0.8}$& 1.3~$\err{0.3}{0.5}$& 
0.47~$\pm0.02$ & 0.97~$\err{0.07}{0.08}$        \\
$n_{\rm e}t~(10^{10}{\rm cm}^{-3}{\rm s})$ &
5.8~$\err{1.2}{1.4}$ & \nodata & \nodata & \nodata & \nodata \\
\\[0mm]

Neon Line: & & & & &    \\
$I_{\rm Ne}$~($10^{-3}$ ph cm$^{-2}$s$^{-1}$) \tablenotemark{c} &
\nodata & \nodata & 
7.9~$\err{4.5}{5.2}$ & 0.83~$\err{0.03}{0.04}$ & 5.5~$\err{1.1}{1.0}$ \\
\\[0mm]

Photoelectric Absorption: & & & & &     \\
$N_{\rm H}~(10^{22}{\rm cm}^{-2})$ &
1.1~$\pm{0.1}$ & 1.2~$\err{0.2}{0.1}$ & 0.84~$\err{0.10}{0.23}$ 
& 0.84~$\fixed$ & 0.84~$\fixed$\\
\\[0mm]
$\chi_{\nu}^2~(\nu)$ &
1.19(26) & 0.80(14) & 0.96(34) & 1.21(40) & 1.25(14)    \\
\enddata
\tablenotetext{a}{Unabsorbed flux (2--10 keV) in units of $10^{-12}\flux$.}
\tablenotetext{b}{Emission Measure: 
$\int n_{\rm e} n_{\rm H}dV / 4\pi D^2$.}
\tablenotetext{c}{Photon flux of herium-like Ne K emission line (0.92 keV).}
\tablecomments{Best-fit values and their 1$\sigma$ errors.}

\end{deluxetable}

\begin{figure}
\figurenum{1}
\epsscale{1.0}
\plottwo{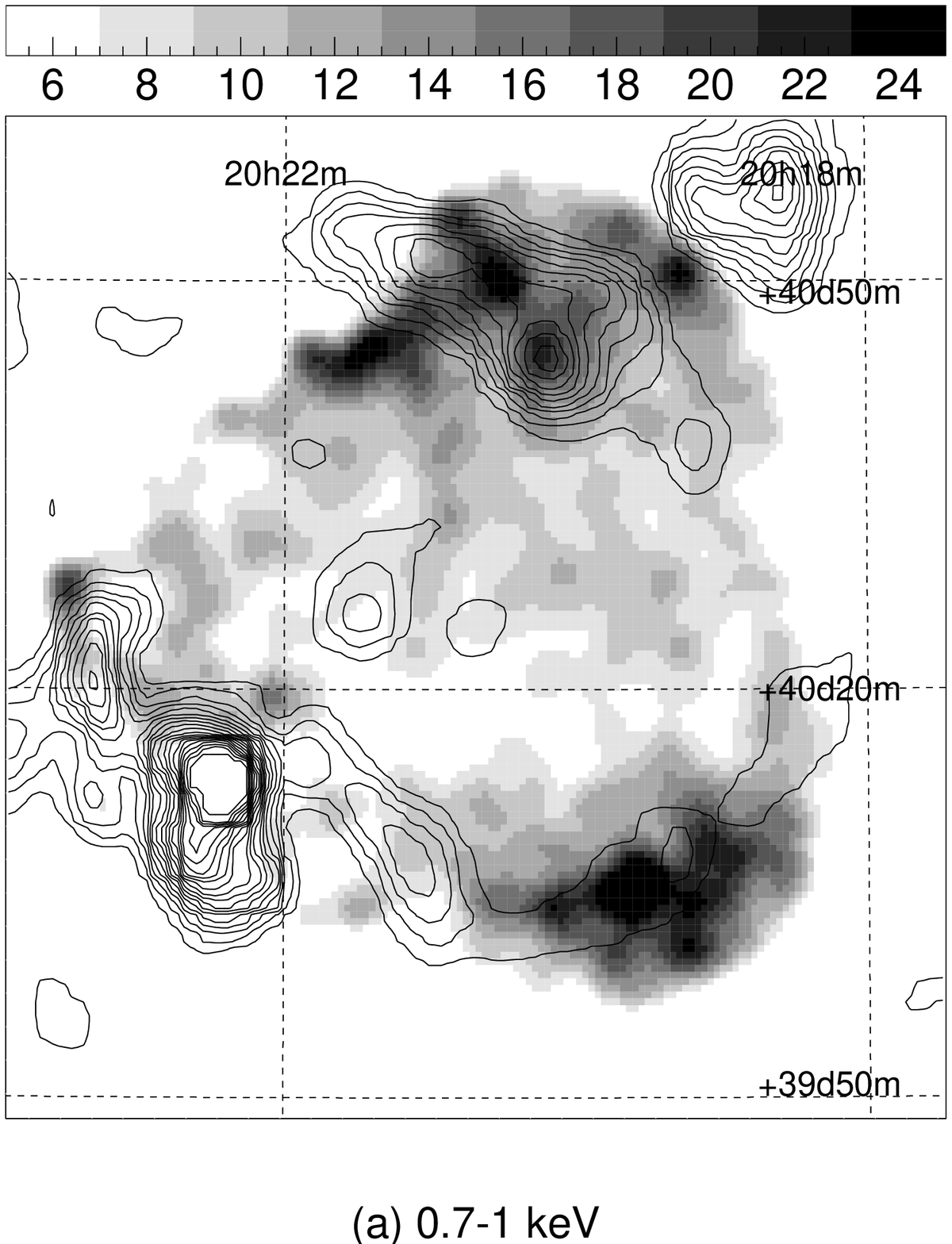}{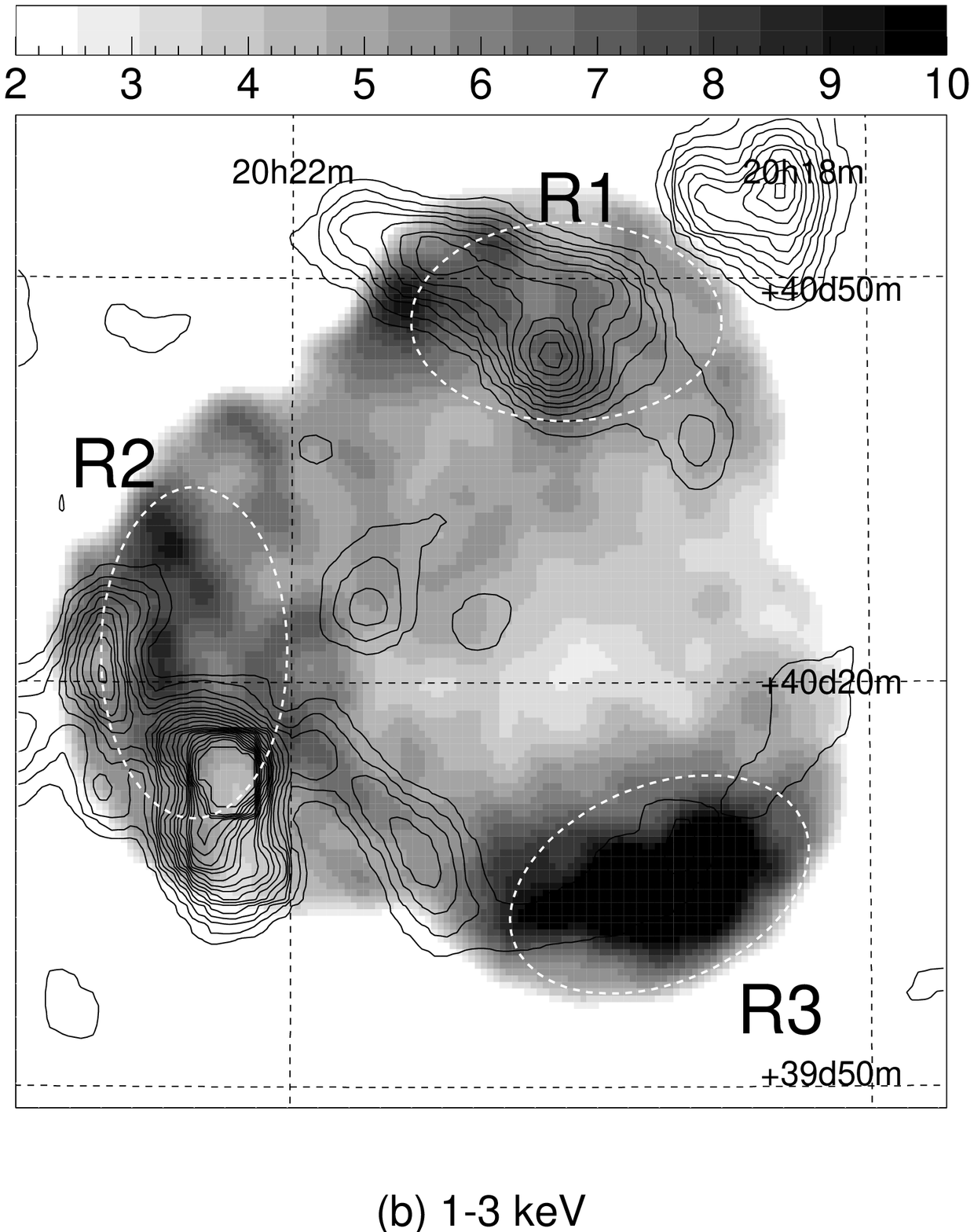}
\caption{GIS images}
\end{figure}

\begin{figure}
\figurenum{1}
\epsscale{0.5}
\plotone{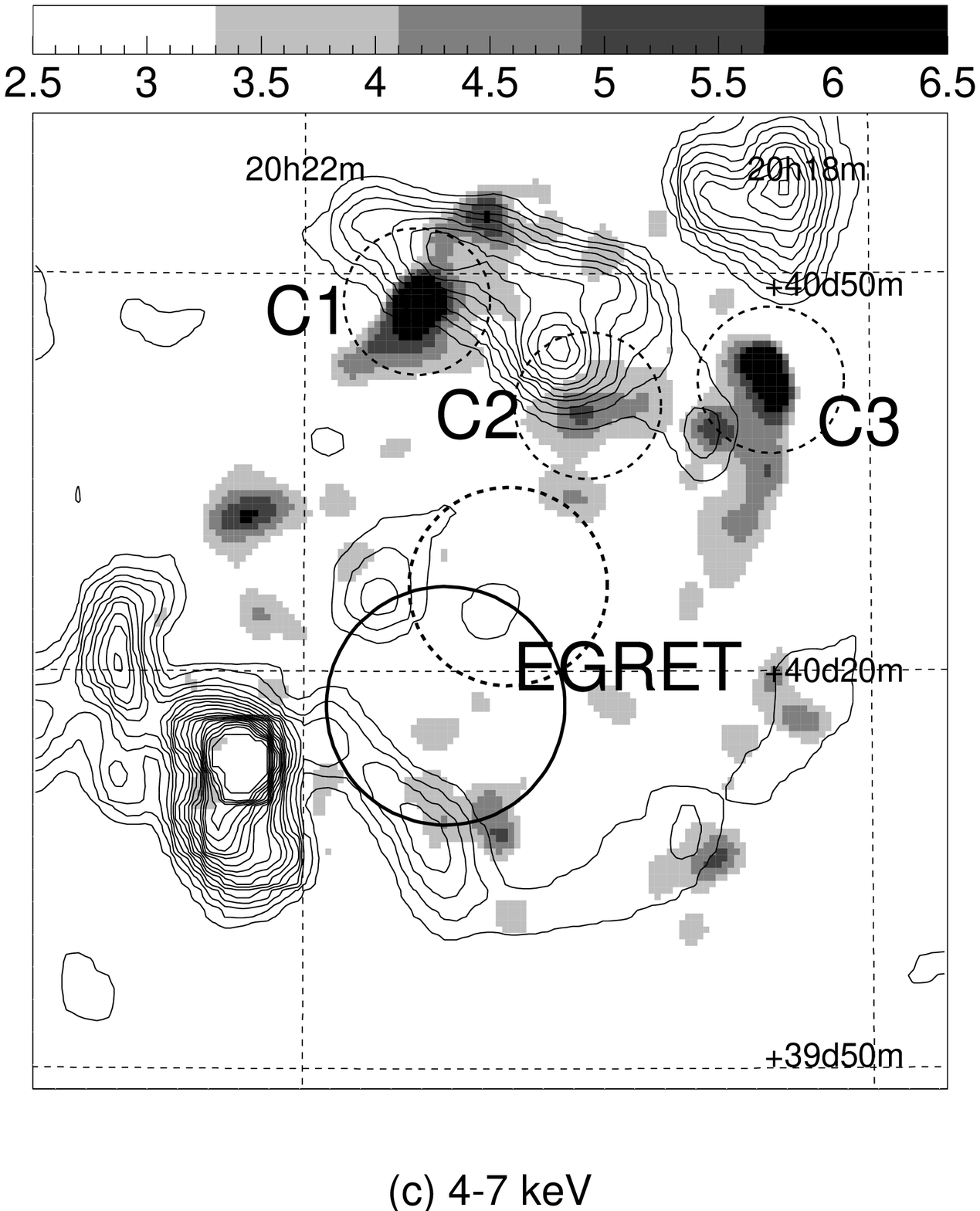}
\caption{GIS images}
\end{figure}

\begin{figure}
\figurenum{2}
\epsscale{0.5}
\plotone{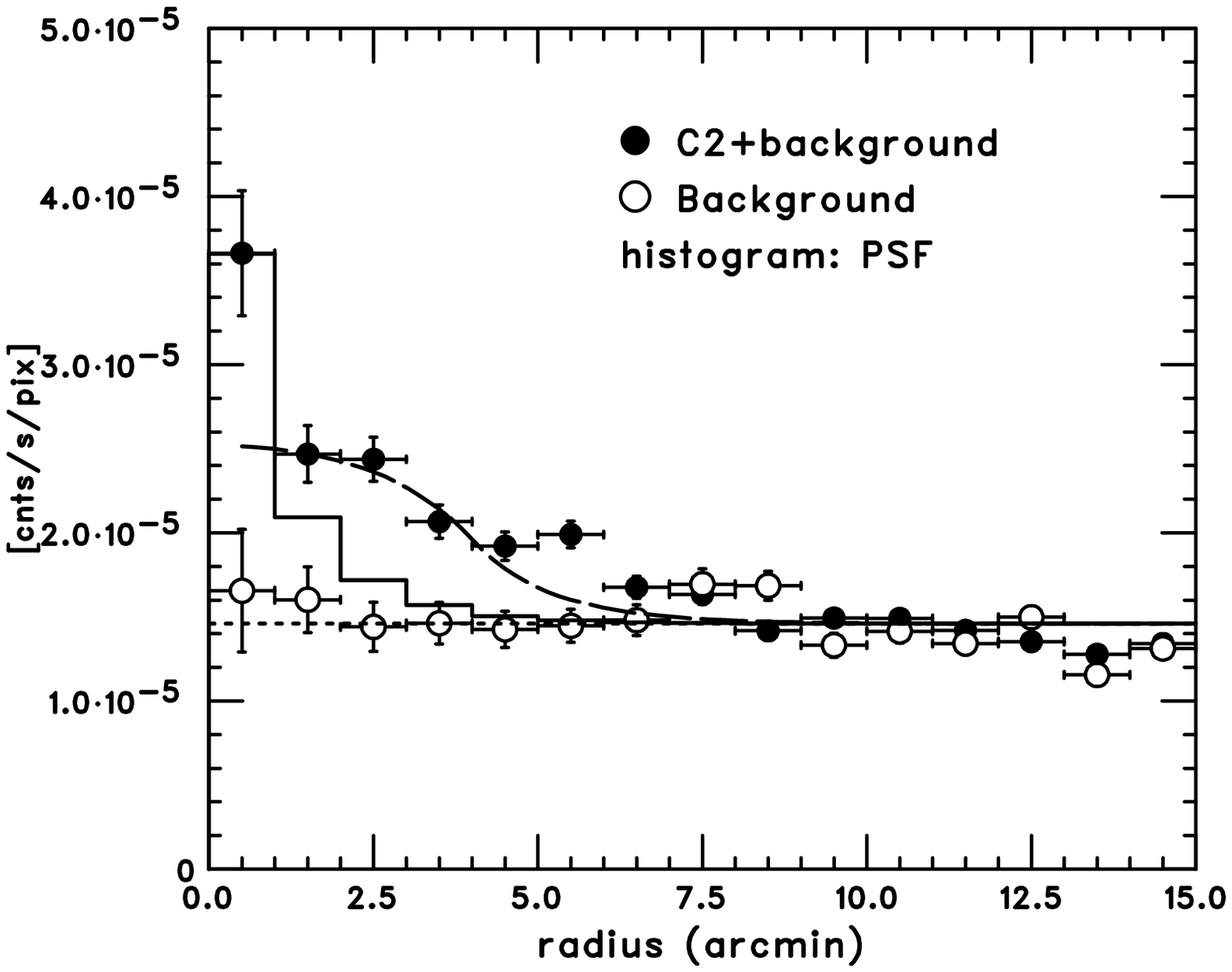}
\caption{Radial Profile of C2}
\end{figure}

\begin{figure}
\figurenum{3}
\epsscale{0.5}
\plotone{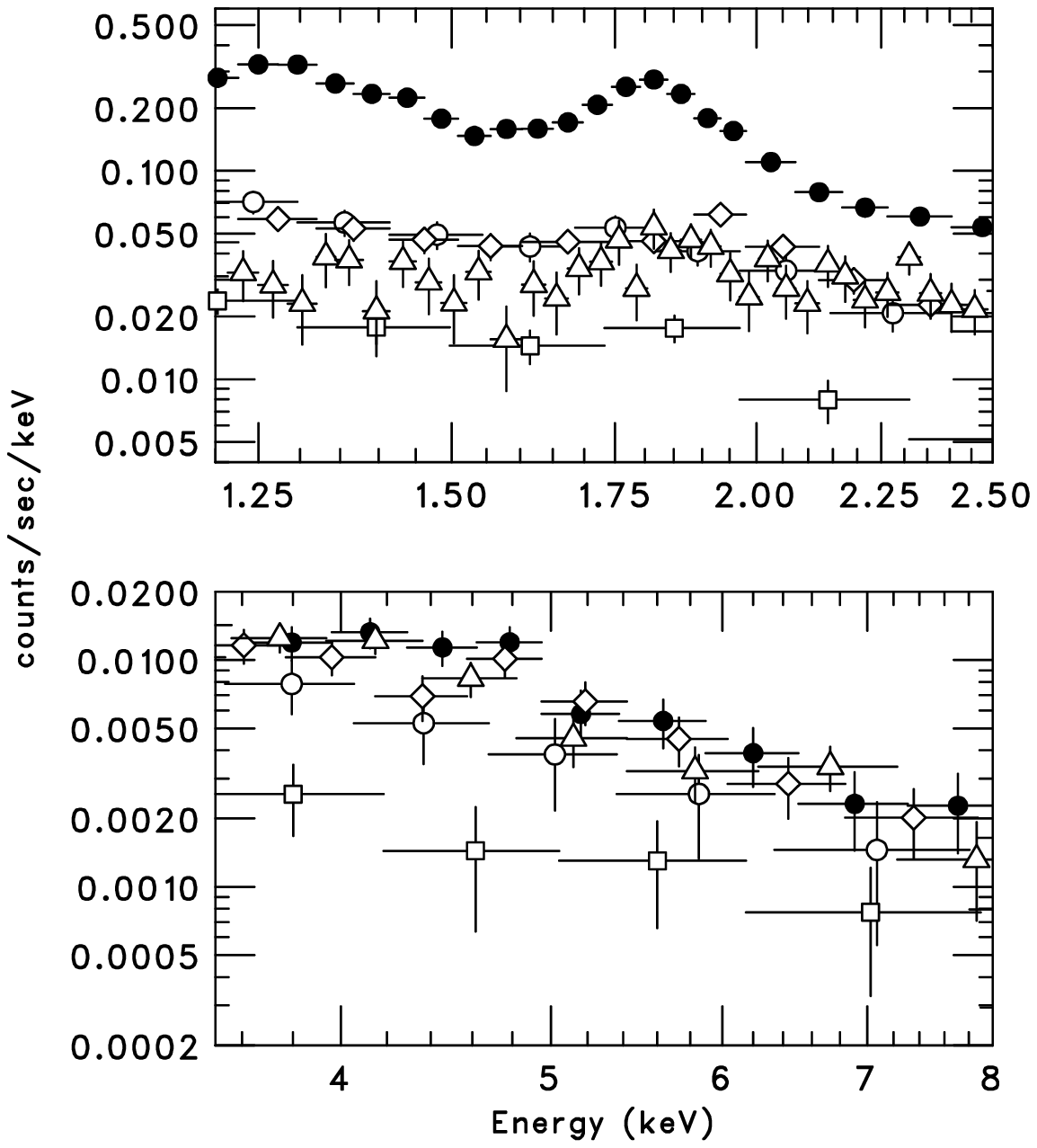}
\caption{Background Fields}
\end{figure}

\begin{figure}
\figurenum{4-5}
\epsscale{1.0}
\plottwo{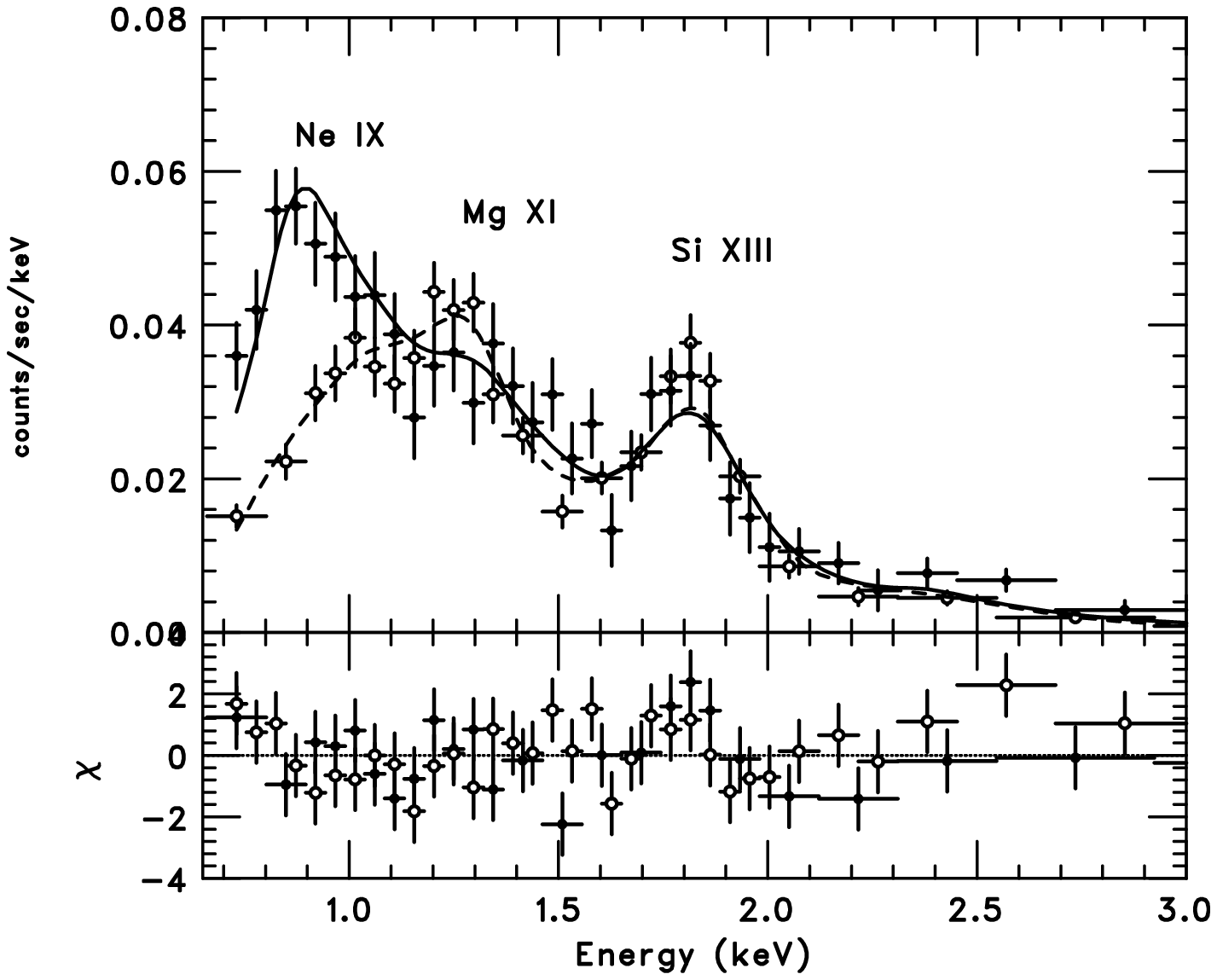}{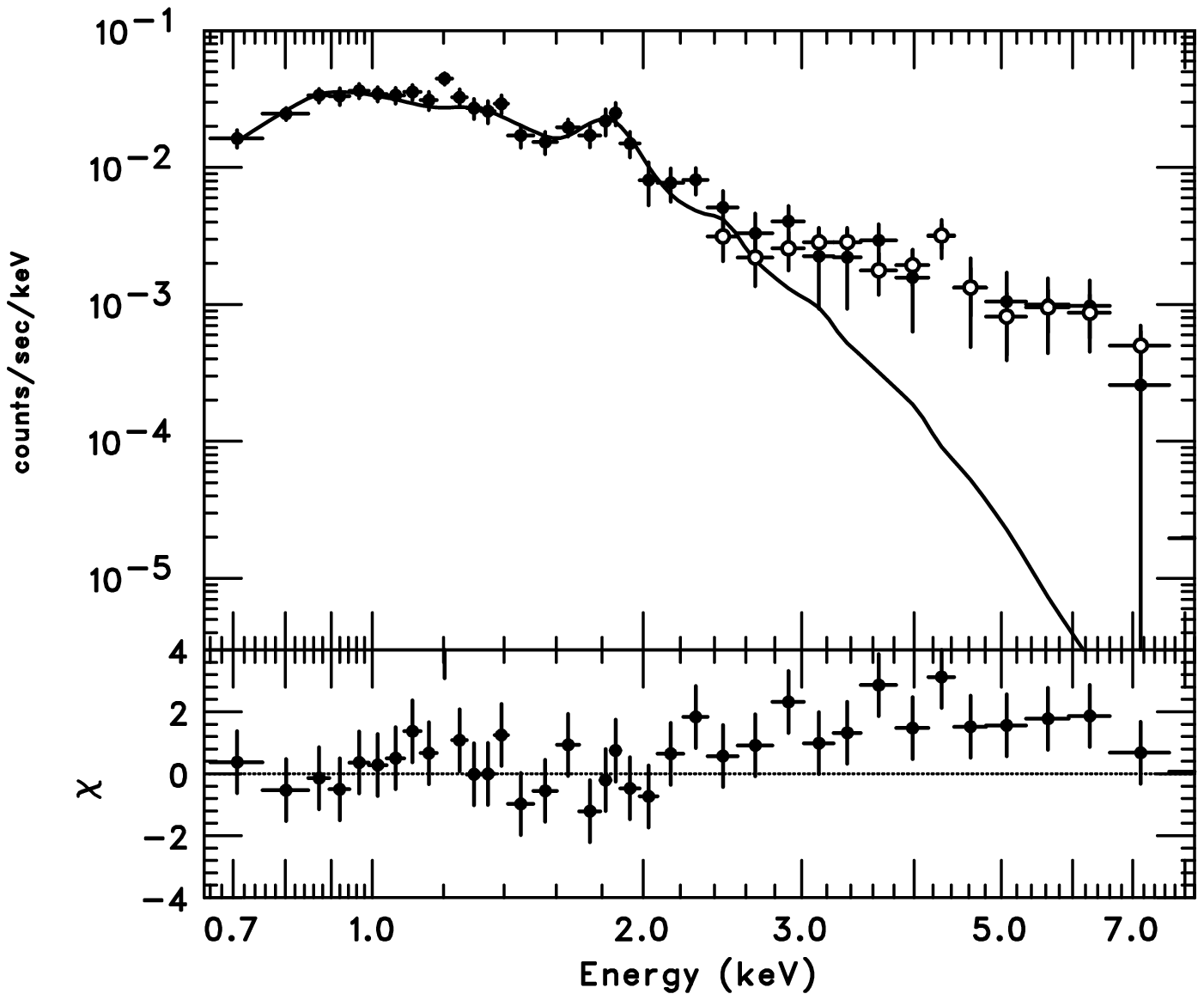}
\caption{GIS spectrum R1/3(left) C2(right)}
\end{figure}

\begin{figure}
\figurenum{6}
\epsscale{0.6}
\plotone{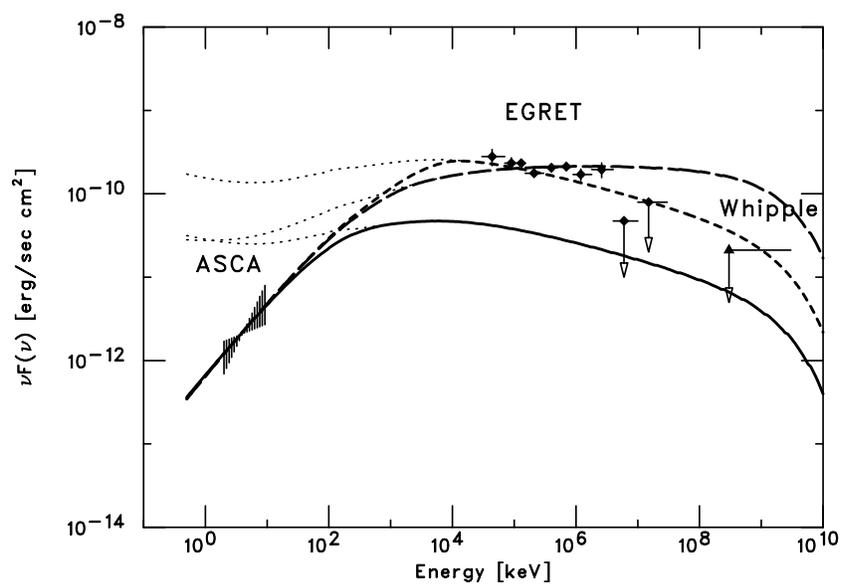}
\caption{Broadband spectrum}
\end{figure}




\begin{thebibliography}{}
\bibitem[e.g. Aharonian, Atoyan, \& Kifune(1997)]{Aharonian97} Aharonian,~F.A.,
        Atoyan,~A.M., \& Kifune,~T.
        1997, \mnras, 291, 162

\bibitem[Aharonian, Drury, \& V\"{o}lk(1994)]{Aharonian94} Aharonian,~F.A.,
        Drury,~L.O'C., \& V\"{o}lk,~H.J.
        1994, \aap, 285, 645

\bibitem[Anders \& Grevesse(1989)]{Angr} Anders,~E., 
        \& Grevesse,~N.
        1989, \gca, 53, 197


\bibitem[Bell(1978)]{Bell78a} Bell, A.R.
	1978, \mnras, 182, 147

\bibitem[Blandford \& Eichler(1987)]{BlandfordEichler87} Blandford, R.
        \& Eichler, D.
        1987, \physrep, 154, 1

\bibitem[see e.g. Blumenthal \& Gould(1970)]{Blumenthal} Blumenthal,~G.R.,
        \& Gould,~R.J.
        1970, Rev. Mod. Phys., 42(2), 237

\bibitem[Brazier et al.(1996)]{Brazier} Brazier,~K.T.S., 
        Kanbach,~G., Carraminana,~A., Guichard,~J., \& Merck,~M.
        1996, \mnras, 281, 1033

\bibitem[Buckley et al.(1998)]{Buckley} Buckley,~J.H., et al.
        1998, \aap, 329, 639

\bibitem[Buote(1999)]{Buote} Buote,~D.A.
        1999, \mnras, 309, 685

\bibitem[SIS; Burke et al.(1994)]{Bur94} Burke,~E.B., 
        Mountain,~R.W., Daniels,~P.J., Cooper,~M.J., and Dorat,~V.S.
        1994, \ieee, 41, 375

\bibitem[Bykov et al.(2000)]{Bykov} Bykov,~A.M.,
        Chevalier,~R.A., Ellison,~D.C., \& Uvarov,~Yu.A.
        2000, \apj, 538, 203

\bibitem[Chevalier(1999)]{Chevalier99} Chevalier,~R.A.
        1999, \apj, 511, 798

\bibitem[Combi, Romero, \& Benaglia(1998)]{combi98} Combi,~J.A.,
        Romero,~G.E., \& Benaglia,~P.
        1998, \aap, 333, L91

\bibitem[Combi et al.(2001)]{combi01} Combi,~J.A.,
        Romero,~G.E., Benaglia,~P., \& Jonas,~J.L.
        2001, \aap, 366, 1047

\bibitem[Downes \& Rinehart(1966)]{DR} Downes,~D., 
        \& Rinehart,~R.
        1966, \apj, 144, 937

\bibitem[Drury, Aharonian, \& V\"{o}lk(1994)]{Drury94} Drury,~L.O'C.,
        Aharonian,~F.A., \& V\"{o}lk,~H.J.
        1994, \aap, 287, 959


\bibitem[Esposito et al.(1996)]{Esposito} Esposito,~J.A.,
        Hunter,~S.D., Kanbach,~G., \& Sreekumar,~P.
        1996, \apj, 461, 820

\bibitem[Gaisser, Protheroe, \& Stanev(1998)]{Gaisser98} Gaisser,~T.K.,
        Protheroe,~R.J., \& Stanev,~T.
        1998, \apj, 429, 219

\bibitem[Green(2000)]{Green} Green,~D.A.
        ``A Catalogue of Galactic Supernova Remnants (2000 August
        version)'', Mullard Radio Astronomy Observatory,
        Cavendish Laboratory, Cambridge, UK

\bibitem[Hartman et al.(1999)]{ThirdEGRET} Hartman,~R.C., et al.
        1999, \apjs, 123, 79


\bibitem[see e.g. Hayakawa (1969)]{Hayakawa69} Hayakawa,~S.
        1969, Cosmic Ray Physics, Wiley-Interscience

\bibitem[Higgs et al.(1977)]{Higgs77} Higgs,~L.A., 
        Landecker,~T.L., \& Roger,~R.S.
        1977, \aj, 82, 718

\bibitem[Hunter et al.(1997)]{Hunter97} Hunter,~S.D., et al.
        1997, \apj, 481, 205

\bibitem[NEI; Itoh(1979)]{Itoh} Itoh,~H.
        1979, \pasj, 31, 541

\bibitem[Jones \& Kang(1993)]{Jones93} Jones,~T.W.,
        \& Kang,~H.
        1993, \apj, 402, 560

\bibitem[Landecker, Roger \& Higgs(1980)]{Land80} Landecker,~T.L., 
        Roger,~R.S., \& Higgs,~L.A.
        1980, \aaps, 39, 133

\bibitem[Liedahl et al.(1990)]{Liedahl90} Liedahl,~D.A,
        Kahn,~S.M., Osterheld,~A.L., \& Goldstein,~W.H.
        1990, \apjl, 350, 37

\bibitem[Maeda et al.(1999)]{Maeda} Maeda,~Y.,
        Koyama,~K., Yokogawa,~J., \& Skinner,~S.
        1999, \apj, 510, 967
        
\bibitem[GIS; Makishima et al.(1996)]{Max96} Makishima,~K., et al.
        1996, \pasj, 48, 171

\bibitem[Masai(1994)]{Masai} Masai,~K.
        1994, \apj, 437, 770

\bibitem[Mewe, Gronenschild, \& Oord(1985)]{Mewe85} Mewe,~R.,
        Gronenschild,~E.H.B.M., \& van~den~Oord,~G.H.J.
        1985, \aaps, 62, 197

\bibitem[Morrison \& McCammon(1983)]{wabs} Morrison,~R.,
        McCammon,~D.
        1983, \apj, 270, 119

\bibitem[e.g., Naito et al.(1999)]{Naito99} Naito,~T.,
        Yoshida,~T., Mori,~M., \& Tanimori,~T.
        1999, Astron. Nachr., 320, 205

\bibitem[Pollock(1985)]{Pollock} Pollock,~A.M.T.,
        1985, \aap, 150, 339

\bibitem[Rephaeli(1979)]{Rephaeli} Rephaeli,~Y.,
        1979, \apj, 227, 364

\bibitem[Saken, Fesen, \& Shull(1992)]{Saken} Saken,~J.M.,
        Fesen, L.A., \& Shull, M.
        1992, \apjs, 81, 715


\bibitem[Sturner \& Dermer(1995)]{Sturner95} Sturner,~S.J.
        \& Dermer,~C.D.
        1995, \aap, 293, L17

\bibitem[Thompson et al.(1995)]{SecondEGRET} Thompson,~D.J., et al.
        1995, \apjs, 101, 259


\bibitem[Wendker, Higgs, \& Landecker(1991)]{Wendker91} Wendker,~H.J.,
        Higgs,~L.A., \& Landecker,~T.L.
        1991, \aap, 241, 551

\bibitem[White \& Long(1991)]{WhiteLong} White,~R.L.,
        \& Long,~K.S.
        1991, \apj, 373, 543

\bibitem[Zhang et al.(1997)]{Zhang} Zhang,~X.,
        Zheng,~Y., Landecker,~T.L., \& Higgs,~L.A.
        1997, \aap, 324, 641

\end{thebibliography}
\end{document}